\documentclass[aps,pra,preprint,superscriptaddress,showpacs]{revtex4}
\usepackage{bm}
\usepackage{amsmath}
\usepackage{amssymb}
\usepackage{mathrsfs}
\usepackage[dvips]{graphicx}
\begin{document}
\title{Polarization of the electron and positron produced in combined Coulomb and strong laser fields}

\author{A. Di Piazza}
\email{dipiazza@mpi-hd.mpg.de}
\affiliation{Max-Planck-Institut f\"ur Kernphysik, Saupfercheckweg 1, 69117 Heidelberg, Germany}

\author{A. I. Milstein}
\email{milstein@inp.nsk.su}
\affiliation{Max-Planck-Institut f\"ur Kernphysik, Saupfercheckweg 1, 69117 Heidelberg, Germany}
\affiliation{Budker Institute of Nuclear Physics, 630090 Novosibirsk, Russia}

\author{C. M\"uller}
\email{c.mueller@mpi-hd.mpg.de}
\affiliation{Max-Planck-Institut f\"ur Kernphysik, Saupfercheckweg 1, 69117 Heidelberg, Germany}

\date{\today}

\begin{abstract}
The process of $e^+e^-$ production in the superposition of a Coulomb and a strong laser field is considered. The pair production rate integrated over the momentum and summed over the spin projections of one of the particles is derived exactly in the parameters of the laser field and in the Born approximation with respect to the Coulomb field. The case of a monochromatic circularly polarized laser field is considered in detail. A very compact analytical expression of the pair production rate and its dependence on the polarization of one of the created particles is obtained in the quasiclassical approximation for the experimentally relevant case of an undercritical laser field. As a result, the polarization of the created electron (positron) is derived.
\end{abstract}

\pacs{12.20.Ds, 42.62.-b}

\maketitle

\section{Introduction}
The creation of electron-positron ($e^+e^-$) pairs in very strong laser fields was studied by theoreticians already in the 1960s and 1970s. They considered pair creation by a high-energy non-laser photon propagating in a strong laser field \cite{Reiss,Nikishov}, by a nuclear Coulomb field in the presence of a strong laser field \cite{Yakovlev}, and by two counterpropagating laser beams forming a standing light wave \cite{Brezin,Popov}. Note that an additional field is always required to induce the pair creation, since a single electromagnetic plane wave does not polarize the vacuum \cite{Schwinger}.

The interest in laser-induced pair creation processes has been strongly revived in recent years due to a large increase in the available laser intensities and, particularly, due to the experiment E-144 conducted at SLAC \cite{SLAC} which achieved the first realization ever of pair creation in laser fields.  Operating in the weakly nonperturbative regime \cite{Huayu}, the experiment observed pair creation by a high-energy $\gamma$ photon of energy $\omega_\gamma\approx 30$\,GeV counterpropagating an intense optical laser pulse. The pairs were produced via the reaction $\omega_\gamma + n\omega_0\to e^+e^-$ with participation of $n\approx 5$ laser photons of energy $\omega_0\approx 2.3$\,eV. This nonlinear process may be considered as the generalization of the Breit-Wheeler process \cite{BW} to intense photon fields. The high-energy $\gamma$ photon was generated experimentally via Compton backscattering of a laser photon off the 46 GeV electron beam from SLAC's linear accelerator.

In the time after the SLAC E-144 experiment several theory groups have studied pair creation in highly energetic laser-particle collisions. Motivated by the advent of the powerful ion accelerator LHC at CERN, the focus has mostly been laid on pair creation by relativistic nuclei colliding with intense laser beams. Here the strong-field variant of the Bethe-Heitler process \cite{BH} can be probed: $Z + n\omega_0\to Z + e^+e^-$, with the nuclear charge number $Z$. Total production rates have been obtained in various interaction regimes \cite{MVG,Avetissian,Sieczka,Milstein}. Energy spectra and angular distributions of the produced particles were calculated \cite{Sieczka,MVG,KuchievPRA,KaisaCorr}. The influence of an additional high-frequency photon \cite{ADP}, of bound atomic states \cite{bf}, and of the nuclear recoil \cite{Sarah,KaisaRecoil} have been studied as well. We note that pair production in counterpropagating laser beams has also been under active scrutiny (see e.g. \cite{Alk,Av,Di_Piazza,Heb,Ruf,Mock,Bul}).

One aspect of the strong-field Bethe-Heitler process, which so far has not been addressed yet,
are the polarization properties of the produced particles. From the ordinary Bethe-Heitler effect (i.e. the conversion of a single high-energy photon into an $e^+e^-$ pair in the field of an atomic nucleus) it is known that it can lead to significant degrees of polarization when the photon energy is high \cite{beam}. The first theoretical studies of this subject date back to the late 1950s \cite{McVoy}. It is an interesting question how the electron and positron spins are oriented after having been extracted from the vacuum by the absorption of not a single photon, but of several or even many photons from a strong laser field. A partial answer to this question has been provided only very recently based on a consideration of the helicity amplitudes of the strong-field Bethe-Heitler process in a circularly polarized laser wave \cite{TimOliver}. It was found that the resulting degree of longitudinal polarization (i.e. the average helicity of the particles) is rather low. Moreover, in \cite{TimOliver} a comparison was drawn between the pair production of fermionic versus bosonic particles.

In this paper, we give a more general treatment of spin effects in the strong-field Bethe-Heitler process by calculating the intrinsic polarization vector of one of the produced particles. First, an expression for the rate, differential in the momentum of one of the particles, is derived which is valid for arbitrary laser field strengths and frequencies. The case of a monochromatic, circularly polarized laser field is investigated in detail. Special emphasis then is placed on the quasiclassical regime of the process, where the laser frequency (in the nuclear rest frame) is small, $\omega\ll m$, and the value of the Lorentz-invariant laser intensity parameter is high, $\xi = |e|E/m\omega\gg 1$. Here, $e$ and $m$ denote the electron charge and mass, respectively, and we use  units with $\hbar = c = 1$. A compact formula for the polarization is obtained in this domain for undercritical laser fields, i.e. at $E\ll E_{cr}$ (in the nuclear rest frame), where $E_{cr}=m^2/|e|=1.3\times 10^{16}\;\text{V/cm}$ is the critical field of quantum electrodynamics. Note that in this regime of laser parameters a very large number of laser photons (typically $n\sim (m/\omega)\xi^2$) participates in the process. Our calculation shows a way how to avoid the summation over the photon number $n$, which arises in the usual treatment of the process and whose evaluation represents a tremendous task in the highly nonperturbative domain ($\xi\gg 1$).

We note that effects of the electron spin degree of freedom have been studied before with regard to various processes occurring in strong laser fields (see also \cite{reviews} for recent reviews). In particular, spin effects were studied in the strong-field Breit-Wheeler process, where a consideration based on helicity amplitudes was provided \cite{Tsai} and the polarization vectors of the created particles were calculated \cite{Serbo}. In both investigations the laser intensity parameter was restricted to moderate values, $\xi\le 1$. Differences between fermionic and bosonic particles have been revealed in pair creation in a standing laser wave \cite{Popov2} and in recent numerical investigations of the Klein paradox \cite{Grobe}. The impact of spin effects was also investigated with respect to strongly laser-driven electron dynamics \cite{Walser}, laser-assisted potential scattering \cite{Mott}, nonlinear Compton scattering \cite{Tsai,Compton} (see also the monograph \cite{Av_book} for further references) and strong-field photoionization \cite{Faisal}.

Our paper is organized as follows. In section II the general expression of the pair production probability integrated over the quantum numbers of one of the created particles is obtained exactly in the laser field parameters and in Born approximation in the Coulomb field. In section III the case of a monochromatic circularly polarized laser field is investigated in detail. In section IV an analytical expression of the pair production rate is calculated in the quasiclassical regime and for undercritical laser fields. The conclusions are presented in section V.

\section{General discussion}
We investigate the process of $e^+e^-$ pair production in the Coulomb field of a nucleus with charge number $Z$ and in a laser field, represented by a plane wave with four-vector potential $A^{\mu}(\phi)=(0,\bm{A}(\phi))$, with $\phi=\kappa x$ being the laser phase, $\kappa^2=0$, $\bm{\kappa}\bm{A}(\phi)=0$ and $\bm{\kappa}$ being the unit vector along the $z$ direction. All the calculations are performed in the rest frame of the nucleus. The S-matrix element of the process in the Born approximation in the Coulomb field and exactly in the parameters of the laser is given by
\begin{equation}
\label{Amp}
S_{fi}=-i\int d^4x \frac{Z\alpha}{r} \bar{U}_{p_1,\lambda_1}(x)\gamma^0 V_{p_2,\lambda_2}(x),
\end{equation}
where $\alpha=e^2$ is the fine-structure constant, $p_1^{\mu}=(\epsilon_1,\bm{p}_1)$ ($p_2^{\mu}=(\epsilon_2,\bm{p}_2)$) with $\epsilon_1=\sqrt{m^2+\bm{p}_1^2}$ ($\epsilon_2=\sqrt{m^2+\bm{p}_2^2}$) is the four-momentum of the created electron (positron), $\lambda_1$ ($\lambda_2$) is its spin quantum number. The wave functions in Eq. (\ref{Amp}),
\begin{eqnarray}\label{eq:ww}
&&U_{p,\lambda}(x)=\mbox{e}^{-ipx}F_p(\phi)R_p(\phi)u_{p,\lambda}\, ,\quad
V_{p,\lambda}(x)=\mbox{e}^{ipx}F_{-p}(\phi)R_{-p}(\phi)v_{p,\lambda}
\end{eqnarray}
are the positive- and negative-energy Volkov states, respectively, with
\begin{eqnarray}
&&F_p(\phi)=\exp\left[-\frac{i}{\kappa p}\int\limits_0^\phi d\eta\left(epA(\eta)-\frac{1}{2}e^2A^2(\eta)\right)\right]\,,\quad
R_p(\phi)=1+\frac{e\hat\kappa\hat A(\phi)}{2\kappa p}\, .
\end{eqnarray}
The spinors $u_{p,\lambda} $ and $v_{p,\lambda}$ in Eq. (\ref{eq:ww}) are free constant Dirac spinors \cite{Landau_b_4_1982} normalized as $u^{\dag}_{p,\lambda}u_{p,\lambda}=v^{\dag}_{p,\lambda}v_{p,\lambda}=1$.

The general expression of the differential pair production probability can be written as
\begin{eqnarray}\label{eq:general}
dW&=&\frac{d\bm p_1 d\bm p_2}{(2\pi)^6}\left|S_{fi}\right|^2=(4\pi Z\alpha)^2\frac{d\bm p_1 d\bm p_2}{(2\pi)^6}\int\!\! \int  d^4xd^4x'\int\!\!\int \frac{d\bm Q d\bm Q'}{(2\pi)^6Q^2 Q'^2}
\nonumber\\
&&\times\mbox{e}^{i\bm Q\bm x} \mbox{e}^{-i\bm Q'\bm x'} \bar{U}_{p_1,\lambda_1}(x)\gamma^0 V_{p_2,\lambda_2}(x)
\bar{V}_{p_2,\lambda_2}(x')\gamma^0 U_{p_1,\lambda_1}(x'),
\end{eqnarray}
where we have written the Coulomb potential in momentum space. It is convenient to pass from the variables $t$ and $z=\bm{\kappa}\bm{x}$ ($t'$ and $z'=\bm{\kappa}\bm{x}'$) to the variables $L=(t+z)/2$ and $\phi=t-z$ ($L'=(t'+z')/2$ and $\phi'=t'-z'$) because the integrals over $L$, $L'$, $\bm x_\perp=\bm{x}-(\bm{\kappa}\bm{x})\bm{\kappa}$, $\bm x_\perp'=\bm{x}'-(\bm{\kappa}\bm{x}')\bm{\kappa}$, and $\bm Q'$ can be performed exactly. As a result we obtain
\begin{eqnarray}\label{eq:general1}
dW&=&(4\pi Z\alpha)^2\frac{d\bm p_1 d\bm p_2}{(2\pi)^3}\int\frac{d\bm Q}{(2\pi)^3Q^4}
\delta(\kappa p_1+\kappa p_2 -\kappa Q)\delta(\bm p_{1\perp}+\bm p_{2\perp}-\bm Q_\perp)
\nonumber\\
&&\times\int\!\! \int  d\phi d\phi'\mbox{e}^{i(\phi-\phi')(\epsilon_1+\epsilon_2)}F_{p_1}^*(\phi)F_{-p_2}(\phi) F_{p_1}(\phi')F_{-p_2}^*(\phi')\nonumber\\
&&\times\bar{u}_{p_1,\lambda_1}R_{-p_1}(\phi)\gamma^0R_{-p_2}(\phi) v_{p_2,\lambda_2}\bar v_{p_2,\lambda_2}        R_{p_2}(\phi')\gamma^0R_{p_1}(\phi')u_{p_1,\lambda_1}\, .
\end{eqnarray}
We want to obtain the dependence of the probability on the quantum numbers of one of the created particles, for example the electron. Therefore, we integrate over the momentum $\bm{p}_2$ of the positron and sum over its spin projections. The differential probability for the positron can be obtained from that of the electron by means of the substitutions $Z\to -Z$ and $e\to -e$. Then, by employing the relation 
\begin{equation}
\sum_{\lambda_2}  v_{p_2,\lambda_2}\bar v_{p_2,\lambda_2}=\frac{1}{2\epsilon_2}(\gamma^0\epsilon_2 -\bm\gamma\bm p_2-m)\,,
\end{equation}
and by integrating over the vector $\bm p_2$, we have
\begin{eqnarray}\label{eq:general2}
dW&=&(4\pi Z\alpha)^2\frac{d\bm p}{4(2\pi)^3}\int\frac{d\bm Q}{(2\pi)^3Q^4}\int\!\! \int  d\phi d\phi'\mbox{e}^{i(\phi-\phi')\Omega_0}
\nonumber\\
&&\times F_{p}^*(\phi)F_{p-Q}(\phi) F_{p}(\phi')F_{p-Q}^*(\phi')\,\vartheta(\kappa Q-\kappa p)\nonumber\\
&&\times\bar{u}_{p,\lambda}R_{-p}(\phi)\gamma^0R_{p-Q}(\phi)\
 R_{Q-p}(\phi')\gamma^0R_{p}(\phi')u_{p,\lambda}\, ,\nonumber\\
&&{ M}=\gamma^0+\gamma^3+\hat\kappa\frac{m^2+(\bm Q_\perp-\bm p_{\perp})^2}{(\kappa,Q-p)^2}-
2\frac{m+\bm\gamma(\bm Q_\perp-\bm p_{\perp})}{(\kappa,Q-p)}\, ,\nonumber\\
&&\Omega_0=\frac{(\kappa p)^2+ p_{\perp}^2+m^2}{2(\kappa p)}+\frac{(\kappa,Q-p)^2+
(\bm Q_\perp-\bm p_{\perp})^2+m^2}{2(\kappa,Q-p)}\, ,
\end{eqnarray}
where $\vartheta(x)$ is the step-function and where for notational simplicity we have removed the subscript $1$ from the quantities referring to the electron. The above expression of the creation probability is valid for a plane wave field with arbitrary shape and polarization.

\section{The case of a monochromatic circularly polarized plane wave}
In order to  investigate the electron polarization, we consider the case of a monochromatic laser field with circular polarization:
\begin{equation}
\bm A(\phi)=\frac{a}{e}\Big(\bm e_1 \cos(\omega\phi)+\mu\bm e_2\sin(\omega\phi)\Big) \,,\quad \mu=\pm1\,,
\end{equation}
where $\bm e_1$ and $\bm e_2$ are two orthogonal unit vectors, perpendicular to $\bm{\kappa}$ and such that $\bm e_1\times\bm e_2=\bm{\kappa}$. Also, $|a|=m\xi$ is a measure of the laser amplitude, $\omega$ is the laser angular frequency and the two values of the parameter $\mu$ correspond to the two possible helicities of the plane wave. It is convenient to introduce the following variables:
\begin{eqnarray}\label{eq:notation}
&&y=\kappa Q-\kappa p\,,\quad y_0=\kappa p\, ,\quad \bm P_\perp=\bm Q_\perp-\bm p_{\perp}\, ,\nonumber\\
&&\bm w =\frac{a}{\omega}\left(\frac{\bm p_{\perp}}{y_0}-\frac{\bm P_\perp}{y} \right)\,,\quad
\sin\psi=\mu \frac{w_2}{w}\, ,\quad \cos\psi=\frac{w_1}{w}\,,\quad w=\sqrt{w_1^2+w_2^2}\,.
\end{eqnarray}
The integrals over the variables $\phi$ and $\phi'$ can be taken and the differential probability results proportional to the laser pulse duration $T_0$. The result for the differential rate $d\dot{W}=dW/T_0$ is
\begin{eqnarray}\label{eq:general3}
d\dot{W}&=&\frac{(4\pi Z\alpha)^2}{4(2\pi)^3}\frac{d\bm p}{(2\pi)^3}\int\limits_0^\infty dy\!\!\int\frac{d\bm P_\perp}{[(y+y_0)^2+
(\bm P_\perp+\bm p_{\perp})^2]^2}
\nonumber\\
&&\times \sum_{n=-\infty}^{\infty}2\pi\delta(\Omega-n\omega)\bar{u}_{p,\lambda}{\bar{\mathcal{T}}}{ M}\mathcal{T} u_{p,\lambda}\, ,\nonumber\\
&&{M}=\gamma^0+\gamma^3+\hat\kappa\frac{m^2+ P_\perp^2}{y^2}-
2\frac{m+\bm\gamma\bm P_\perp}{y}\, ,\nonumber\\
&&\mathcal{T}=\left(\gamma^0-\frac{a^2}{2yy_0}\hat\kappa\right)J_n(w)-\frac{a}{4}\left(\frac{\gamma^0\hat\kappa}{y_0}-
\frac{\hat\kappa\gamma^0}{y}\right)\nonumber\\
&&\times\left[(\gamma^1-i\mu\gamma^2)\mbox{e}^{i\psi}J_{n-1}(w)+(\gamma^1+i\mu\gamma^2)\mbox{e}^{-i\psi}J_{n+1}(w)\right]\,,\nonumber\\ 
&&\bar{\mathcal{T}}=\left(\gamma^0-\frac{a^2}{2yy_0}\hat\kappa\right)J_n(w)-\frac{a}{4}\left(\frac{\hat\kappa\gamma^0}{y_0}-
\frac{\gamma^0\hat\kappa}{y}\right)\nonumber\\
&&\times\left[(\gamma^1+i\mu\gamma^2)\mbox{e}^{-i\psi}J_{n-1}(w)+(\gamma^1-i\mu\gamma^2)\mbox{e}^{i\psi}J_{n+1}(w)\right]\,,\nonumber\\ 
&&\Omega=\frac{y_0^2+ p_{\perp}^2+m^2+a^2}{2y_0}+\frac{y^2+P_\perp^2+m^2+a^2}{2y}\, .
\end{eqnarray}
Note that $\Omega$ is the sum of the electron and positron energy. The quantity ${\cal M}=\bar{u}_{p,\lambda}{\bar{\mathcal{T}}}{ M}\mathcal{T} u_{p,\lambda}$ can be represented as
\begin{eqnarray}
&&{\cal M}=\frac{y_0}{\epsilon}\Big[C_1J_n^2(w)+C_2J_{n-1}^2(w)+C_3J_{n+1}^2(w)\nonumber\\
&&+C_4J_{n-1}(w)J_{n+1}(w)+C_5J_{n}(w)J_{n-1}(w)+C_6J_{n}(w)J_{n+1}(w)\Big]\, ,
\end{eqnarray}
where the coefficients $C_i$, $i=1,\ldots,6$ are given by
\begin{eqnarray}
C_1&=&\left(1-\frac{a^2}{yy_0}\right)^2-2\left(1-\frac{a^2}{yy_0}\right)\frac{(m^2-\bm p_{\perp}\bm P_{\perp})}{yy_0} 
+\frac{(m^2+p_{\perp}^2)(m^2+P_{\perp}^2)}{y^2y_0^2}\, ,\nonumber\\
C_2&=&\frac{a^2}{2y^2y_0^2}\left[p_{\perp}^2+P_{\perp}^2-\mu\frac{(P_{\perp}^2-p_{\perp}^2)}{y_0}
(\bm\zeta,\bm p-\bm F)\right]\,,\nonumber\\
C_3&=&C_2(\mu\to-\mu)\,,\nonumber\\
C_4&=&\frac{2a^2}{y^2y_0^2}\left\{\frac{2[\bm p_{\perp}\times\bm P_{\perp}]^2}{yy_0(\frac{\bm p_{\perp}}{y_0}
-\frac{\bm P_{\perp}}{y})^2}-\bm p_{\perp}\bm P_{\perp}\right\}\,,\nonumber\\
C_5&=&\frac{a}{wyy_0}\Bigg\{\left[ \frac{m^2+p_{\perp}^2}{yy_0}-\left(1-\frac{a^2}{yy_0}\right)\right]\bm P_\perp\bm w
-\left[ \frac{m^2+P_{\perp}^2}{yy_0}-\left(1-\frac{a^2}{yy_0}\right)\right]
\bm p_{\perp}\bm w\nonumber\\
&&+\mu\left[ \frac{m^2+P_{\perp}^2}{yy_0}+\left(1-\frac{a^2}{yy_0}\right)\right]
\bm\zeta\bm G_1\nonumber\\
&&+\frac{\mu}{y_0}\left(1-\frac{a^2}{yy_0}\right)\left[(\bm P_{\perp}\bm w)(\bm\zeta,\bm p-\bm F)
-m\left(\bm\zeta,(\epsilon+m)\bm w-\bm G_1+\bm G_2\right)\right]\nonumber\\
&&+\frac{\mu}{y}\left[(\bm P_{\perp}\bm w)(\bm\zeta,\bm p+\bm F)
-m\left(\bm\zeta,(\epsilon+m)\bm w-\bm G_1-\bm G_2\right)\right]\Bigg\}\,,\nonumber\\
C_6&=&C_5(\mu\to-\mu)\,,\nonumber\\
&&\bm F=m\bm\kappa+\frac{(\bm\kappa\bm p)}{\epsilon+m}\bm p\,,\quad 
\bm G_1=m\bm w+\frac{(\bm w\bm p)}{\epsilon+m}\bm p\,,\quad 
\bm G_2=(\bm p\bm w)\bm\kappa-(\bm p\bm\kappa)\bm w\, .
\end{eqnarray}
Here we introduced the polarization vector $\bm{\zeta}$ in the rest frame of the electron described in the nuclear rest frame by the Dirac spinor $u_{p,\lambda}$ \cite{Landau_b_4_1982}.

Very often the parameters of the laser field correspond to the situation where $\Omega\gg \omega$. In this case it is necessary to take the sum in Eq. (\ref{eq:general3}) over many terms involving very large values of $n$. This is numerically a very hard task. However, this can be avoided in the following way. We set
\begin{eqnarray}
&&B_{1,n}=J_n^2(w)\,,\quad B_{2,n}=J_{n-1}^2(w) \,,\quad B_{3,n}=J_{n+1}^2(w)\,,\nonumber\\
&& B_{4,n}=J_{n-1}(w)J_{n+1}(w)\,,\quad B_{5,n}=J_{n}(w)J_{n-1}(w) \,,\quad B_{6,n}=J_{n}(w)J_{n+1}(w)\,,\nonumber\\
&&S_k=\sum_{n=-\infty}^{\infty}2\pi\delta(\Omega-n\omega)B_{k,n}\,, \quad \text{for } k=1,\ldots,6\,.
\end{eqnarray}
Then, by using the relations \cite{Gradshteyn_b_2000}
\begin{eqnarray}
&& \sum_{n=-\infty}^{\infty}\exp(inb)J_{n}(u)J_{n+k}(v)=\left(\frac{v-u\exp(-ib)}{v-u\exp(ib)}\right)^{k/2}
J_k(\sqrt{u^2+v^2-2uv\cos b})\,,\nonumber\\
&&2\pi\delta(\Omega-n\omega)=4\mbox{Re}\int_0^\infty ds\,\mbox{e}^{2is(n\omega-\Omega)}\, ,
\end{eqnarray}
we obtain
\begin{eqnarray}
&&S_1=4\int_0^\infty ds \cos(2s\Omega)J_0(\eta)\,,\quad S_2=4\int_0^\infty ds \cos[2s(\Omega-\omega)]J_0(\eta)
\,,\nonumber\\
&& S_3=4\int_0^\infty ds \cos[2s(\Omega+\omega)]J_0(\eta)\,,\quad
 S_4=-4\int_0^\infty ds \cos(2s\Omega)J_2(\eta)\,,\nonumber\\
&& S_5=4\int_0^\infty ds \sin[s(2\Omega-\omega)]J_1(\eta) \,,\quad S_6=4\int_0^\infty ds \sin[s(2\Omega+\omega)]J_0(\eta)\,,\nonumber\\
&& \eta=2w\sin(\omega s)\, .
\end{eqnarray}
Finally, it is convenient to use the parametrization, 
\begin{equation}
\label{tau}
\frac{1}{A^2}=-\int_0^\infty d\tau \tau \exp(i\tau A) \,,
\end{equation}
so that the differential rate $d\dot W$ can be written in the form
\begin{eqnarray}\label{eq:general4}
d\dot{W}&=&-\frac{y_0(4\pi Z\alpha)^2}{\epsilon(2\pi)^3}\,\frac{d\bm p}{(2\pi)^3}\,\mbox{Re}\int\limits_0^\infty dy\!\!
\int\limits_0^\infty d\tau\, \tau \int\limits_0^\infty ds\!\!\int\!\! {d\bm P_\perp}\exp(i\Phi_0)
\nonumber\\
&&\times \{[C_1 +(C_2+C_3)\cos(2\omega s)]J_0(\eta)-C_4J_2(\eta)-i(C_5+C_6)\cos(\omega s)J_1(\eta)\nonumber\\
&&-i(C_2-C_3)\sin(2\omega s)J_0(\eta)-(C_5-C_6)\sin(\omega s)J_1(\eta)\}\, ,\nonumber\\
\Phi_0&=&\tau[(y+y_0)^2+(\bm P_\perp+\bm p_{\perp})^2]+2\Omega s\,.
\end{eqnarray}

Thanks to the parametrization in Eq. (\ref{tau}) we can easily take the integral over $\bm P_\perp$. We pass from the variable  $\bm P_\perp$ to the variable
$\bm q=\bm p_{\perp}/y_0-\bm P_\perp/y$ and take the integral first over the angle of the two-dimensional vector $\bm q$ using the independence of the argument of the Bessel functions from this angle. The result is
\begin{eqnarray}\label{eq:general5}
&&d\dot{W}=-\frac{y_0(4\pi Z\alpha)^2}{\epsilon(2\pi)^2}\,\frac{d\bm p}{(2\pi)^3}\,\mbox{Re}\int\limits_0^\infty  dy\, y^2\!\!
\int\limits_0^\infty d\tau\, \tau \int\limits_0^\infty ds\int\limits_0^\infty dq\,q (X+\mu\bm\zeta \bm Y)\exp(i\Phi)\,,
\nonumber\\
&&\Phi=\tau(y+y_0)^2\left(1+\frac{p_{\perp}^2}{y_0^2}\right)+s(y+y_0)\left(1+\frac{ p_{\perp}^2}{y_0^2}+
\frac{m^2+a^2}{yy_0}\right)+y(\tau y+s) q^2\,,\nonumber\\
&&\eta=\frac{2aq}{\omega}\sin(\omega s)\,,\quad \eta_1=2\frac{y}{y_0}qp_{\perp}[\tau (y+y_0)+s]\, , \nonumber
\end{eqnarray}
where the quantities $X$ and $\bm Y$ have the form
\begin{eqnarray}
&&X=\Bigg\{\left(1-\frac{m^2+a^2}{yy_0}+\frac{p_{\perp}^2}{y_0^2}\right)^2+\frac{m^2}{y^2y_0^2}\left(1+\frac{y}{y_0}\right)^2
p_{\perp}^2+\frac{m^2+p_{\perp}^2}{y_0^2}q^2\nonumber\\
&& +\cos(2\omega s) \frac{a^2}{y^2y_0^2}\left[\left(1+\frac{y^2}{y_0^2}\right)p_{\perp}^2+y^2q^2\right]\Bigg\}J_0(\eta_1)J_0(\eta)\nonumber\\
&&+\frac{2iqp_{\perp}}{y_0}\left[1-\frac{a^2}{yy_0}+\frac{m^2+p_{\perp}^2}{y_0^2}+\frac{a^2}{y_0^2}
\cos(2\omega s)\right]J_1(\eta_1)J_0(\eta)\nonumber\\
&&-2\frac{a^2}{y_0^2}\left[\frac{p_{\perp}^2}{yy_0}J_2(\eta_1)-iq\frac{p_{\perp}}{y}J_1(\eta_1)\right]J_2(\eta)\nonumber\\
&&-i\frac{2a}{y_0}\cos(\omega s)\Bigg\{q\left(1-\frac{m^2+p_{\perp}^2+a^2}{yy_0}+  \frac{p_{\perp}^2}{y_0^2}\right)J_0(\eta_1)
-q\frac{p_{\perp}^2}{y_0^2}J_2(\eta_1)\nonumber\\
&&+i\frac{p_{\perp}}{y}\left[\left(1-\frac{y}{y_0}\right)\left(\frac{m^2+a^2}{yy_0}-\frac{p_{\perp}^2}{y_0^2}
-1\right)+\frac{y}{y_0}q^2\right]J_1(\eta_1)\Bigg\}J_1(\eta)\, ,\nonumber
\end{eqnarray}
\begin{eqnarray}
&&\bm Y=i\sin(2\omega s)\frac{a^2}{y_0^3}(\bm p-\bm F)\left\{\left[\left(\frac{1}{y_0^2}-\frac{1}{y^2}\right)p_{\perp}^2+q^2\right]
J_0(\eta_1)+\frac{2i}{y_0}qp_{\perp}J_1(\eta_1)\right\}J_0(\eta)\nonumber\\
&&+\sin(\omega s)\frac{2a}{y y_0^2}\Bigg\{yp_{\perp}\left(\frac{m\bm p_{\perp}}{p_{\perp}^2}+\frac{\bm p}{\epsilon+m}\right)
\left[\left(\frac{p_{\perp}^2}{y_0^2}+q^2\right)iJ_1(\eta_1)+
\frac{qp_{\perp}}{y_0}\left(J_0(\eta_1)-J_2(\eta_1)\right)\right]\nonumber\\
&&+\left(1-\frac{a^2}{yy_0}\right)(\bm p-\bm F)\left[qyJ_0(\eta_1)+ip_{\perp}\left(1+\frac{y}{y_0}\right)J_1(\eta_1)\right]
+(\bm p+\bm F)qy_0J_0(\eta_1)\nonumber\\
&&+ip_{\perp}\left[\frac{m}{y}\Big((\epsilon+m)\bm\kappa-\bm F\Big)+\bm p+\bm F\right]J_1(\eta_1)\Bigg\}J_1(\eta)\,.
\end{eqnarray}
As expected, if we integrate with respect to the electron momentum $\bm{p}$, the resulting pair production rate has the general form $A+\bm{B}\bm{\zeta}$, with the vector $\bm{\zeta}_f=\bm{B}/A$ being the  polarization of the created electron \cite{Landau_b_4_1982}. Also, since the quantity $\bm{\zeta}$ is a pseudo-vector and $\bm{B}$ is a vector, the differential rate $d\dot{W}$ depends on the product $\mu\bm{\zeta}$ (we recall that the values $\mu=\pm 1$ stem from right- and left-handed circular polarization of the plane wave).

Now, the integral over the variable $q$ can be taken by employing the relation \cite{Gradshteyn_b_2000}
\begin{eqnarray}\label{int}
&&\int\limits_0^\infty  dx \mbox{e}^{icx}J_\nu(d\sqrt{x})J_\nu(b\sqrt{x})=
\frac{i}{c}J_\nu\left(\frac{db}{2c}\right)\exp\left(i\frac{\pi\nu}{2}-i\frac{d^2+b^2}{4c}\right)
\,,
\end{eqnarray}
and the recurrence relations of the Bessel functions. The result reads

\begin{eqnarray}\label{eq:general6}
&&d\dot{W}=\frac{(4\pi Z\alpha)^2}{2(2\pi)^2}\,\frac{dy_0\,d\bm p_{\perp}}{(2\pi)^3}\,\mbox{Im}\int\limits_0^\infty  dy\, y\!\!
\int\limits_0^\infty \frac{d\tau\, \tau }{T}\int\limits_0^\infty ds\,s\exp(i\Phi_1)(X_1+\mu\bm\zeta \bm Y_1)\,,\\
&&\Phi_1=sy_0\left[(T+\tau)\left(y+1+\frac{p_{\perp}^2}{y_0^2T}\right)+(y+1)\frac{m^2+a^2}{yy_0^2}-
\frac{a^2\sin^2(\omega s)}{y_0^2(\omega s)^2yT}\right]\,,\nonumber\\
&& T=\tau y+1\,,\quad \rho=\frac{2 a p_{\perp}(T+\tau)\sin(\omega s)}{\omega y_0T}\,,\nonumber
\end{eqnarray}
with
\begin{eqnarray}
&&X_1=\Bigg[\left(1-\frac{m^2+a^2}{yy_0^2}+\frac{p_{\perp}^2}{y_0^2}\right)^2+\frac{m^2p_{\perp}^2}{y^2y_0^4}(1+y)^2
 +\cos(2\omega s) \frac{a^2p_{\perp}^2}{y^2y_0^4}(1+y^2)\Bigg]J_0(\rho)\nonumber\\
&&+\left[\frac{m^2+p_{\perp}^2}{y_0^2}+\cos(2\omega s) \frac{a^2}{y_0^2}\right]
\left\{\left[\frac{a^2\sin^2(\omega s)}{(\omega sy)^2}+p_{\perp}^2(T+\tau)^2\right]\frac{J_0(\rho)}{y_0^2T^2}+
i\frac{J_0(\rho)-\rho J_1(\rho)}{syy_0T}\right\}\nonumber\\
&&-\frac{2p_{\perp}}{y_0^2T}\left[1-\frac{a^2}{yy_0^2}+\frac{m^2+p_{\perp}^2}{y_0^2}+\frac{a^2}{y_0^2}
\cos(2\omega s)\right]\left[p_{\perp}(T+\tau)J_0(\rho) -i\frac{a\sin(\omega s)}{(\omega s)y}J_1(\rho)\right]  \nonumber\\
&&-2\frac{a^2p_{\perp}}{y_0^4yT}\left[p_{\perp}\tau J_2(\rho) +i\frac{a\sin(\omega s)}{(\omega s)y}J_1(\rho)\right]\nonumber\\
&&-i\frac{2a}{y_0}\cos(\omega s)\Bigg\{\frac{1}{Ty_0}\left(1-\frac{m^2+p_{\perp}^2+a^2}{yy_0^2}\right)
\left[p_{\perp}(T+\tau)J_1(\rho) +i\frac{a\sin(\omega s)}{(\omega s)y}J_0(\rho)\right]\nonumber\\
&&+\frac{p_{\perp}^2}{y_0^3T}\left[2p_{\perp}(T+\tau)J_1(\rho) +i\frac{a\sin(\omega s)}{(\omega s)y}[J_0(\rho)
-J_2(\rho)]\right]\nonumber\\
&&-\frac{p_{\perp}}{yy_0}(1-y)\left(\frac{m^2+a^2}{yy_0^2}-\frac{p_{\perp}^2}{y_0^2}
-1\right)J_1(\rho)\, \nonumber\\
&&-\frac{p_{\perp}}{y_0}\left[\left(\frac{a^2\sin^2(\omega s)}{(\omega sy)^2}+p_{\perp}^2(T+\tau)^2\right)\frac{J_1(\rho)}{y_0^2T^2}+
i\frac{\rho J_0(\rho)}{syy_0T}\right]\Bigg\}\,,\nonumber
\end{eqnarray}
\begin{eqnarray}
&&\bm Y_1=i\sin(2\omega s)\frac{a^2}{y_0^3}(\bm p-\bm F)\Bigg\{\left(1-\frac{1}{y^2}\right)\frac{p_{\perp}^2}{y_0^2}
J_0(\rho)\nonumber\\
&&+\left[\frac{a^2\sin^2(\omega s)}{(\omega sy)^2}+p_{\perp}^2(T+\tau)^2\right]\frac{J_0(\rho)}{y_0^2T^2}+
i\frac{J_0(\rho)-\rho J_1(\rho)}{syy_0T}\nonumber\\
&&-\frac{2p_{\perp}}{y_0^2T}\left[p_{\perp}(T+\tau)J_0(\rho) -i\frac{a\sin(\omega s)}{(\omega s)y}J_1(\rho)\right]\Bigg\}  \nonumber\\
&&+\sin(\omega s)\frac{2a}{y y_0^3}\Bigg\{yy_0p_{\perp}\left(\frac{m\bm p_{\perp}}{p_{\perp}^2}+\frac{\bm p}{\epsilon+m}\right)
\Bigg[-\frac{p_{\perp}^2}{y_0^2}J_1(\rho)\nonumber\\
&&- \left(\frac{a^2\sin^2(\omega s)}{(\omega sy)^2}+p_{\perp}^2(T+\tau)^2\right)\frac{J_1(\rho)}{y_0^2T^2}-
i\frac{\rho J_0(\rho)}{syy_0T}\nonumber\\
&&+\frac{p_{\perp}}{y_0^2T}\left[2p_{\perp}(T+\tau)J_1(\rho) +i\frac{a\sin(\omega s)}{(\omega s)y}[J_0(\rho)
-J_2(\rho)]\right]\Bigg]\nonumber\\
&&+\left(1-\frac{a^2}{yy_0^2}\right)(\bm p-\bm F)\Bigg[-p_{\perp}(1+y)J_1(\rho)
+\frac{y}{T}\left[p_{\perp}(T+\tau)J_1(\rho) +i\frac{a\sin(\omega s)}{(\omega s)y}J_0(\rho)\right]\Bigg]\nonumber\\
&&+(\bm p+\bm F)\frac{1}{T}\left[p_{\perp}(T+\tau)J_1(\rho) +i\frac{a\sin(\omega s)}{(\omega s)y}J_0(\rho)\right]\nonumber\\
&&-p_{\perp}\left[\frac{m}{yy_0}\Big((\epsilon+m)\bm\kappa-\bm F\Big)+\bm p+\bm F\right]J_1(\rho)\Bigg\}\,,
\end{eqnarray}
where we also made the replacements $\tau \to s\tau/y_0$ and $y\to y y_0$ and we used the relations
\begin{eqnarray}\label{epsi}
\epsilon= \frac{p_{\perp}^2+m^2}{2y_0}+\frac{y_0}{2}\,,
\quad \bm \kappa\bm p= \frac{p_{\perp}^2+m^2}{2y_0}-\frac{y_0}{2}\,,\quad  
d\bm p=\frac{\epsilon}{y_0}dy_0\,d\bm p_{\perp}\,.
\end{eqnarray} 
So far we have derived the differential rate $d\dot{W}$ for the electron. As it was already pointed out above, the expression for the positron can be obtained from that for the electron by the replacement $Z\to -Z$ and $e\to -e$ (the latter being equivalent to $a\to - a$). Since  the integrand in the formula for $d\dot{W}$ is an even function of $Z$ and $a$, the rate and the average polarization of the positron will be the same as those of the electron. Also, note that in the Born approximation with respect to the Coulomb field $\bm\zeta_f$ is independent of the nuclear charge number $Z$. Below, for convenience, we will assume that $a>0$.  

\section{Quasiclassical limit}
It is of experimental interest to consider the case $\omega/m\ll 1$ and $\xi=a/m\gg 1$ (quasiclassical limit). We also assume that the product $(\omega/m)\xi=E/E_{cr}=\chi$ is much smaller than unity. Note that in order to increase the ratio $E/E_{cr}$ in the rest frame of the nucleus it is possible to use ultra-relativistic nuclear beams counterpropagating the laser wave. In this case, $E/E_{cr}\approx 2\gamma E_0/E_{cr}$, where $\gamma$ is the nuclear gamma factor and where $E_0$ is the laser field strength in the laboratory frame, which can be of the order of $10^{-5}E_{cr}$ for table-top optical lasers (corresponding to values of $\xi$ of the order of $10$) \cite{Lasers}. 

In the quasiclassical limit, the phase $\Phi_1$ is large. Also, the argument $\rho$ of the Bessel functions is large in the region of parameters which gives the biggest contribution to the rate (we will show this below). Since $\Phi_1$ is positive, in order to get compensation in the total phase, we can make the substitution  \cite{Gradshteyn_b_2000}
\begin{equation}
J_n(\rho)\to \frac{1}{\sqrt{2\pi\rho}}\exp\left(-i\rho+i\frac{\pi}{4}+i\frac{\pi n}{2}\right)\left[1-\frac{i}{2\rho}\left(n^2-\frac{1}{4}\right)\right]\,,
\end{equation}
using the asymptotic of the Bessel functions. Note that the next-to-leading order term in $\rho$ in the pre-exponent of the Bessel functions has also to be taken in order to obtain the correct leading contribution to the spin-dependent part of the rate. Also, we will see that $\omega s$ is a small quantity in the region of parameters which gives the biggest contribution to the rate, and this allows one to expand with respect to $\omega s$. In this way we obtain the following expression for the total phase $\Phi_2=\Phi_1-\rho$:
\begin{eqnarray}
\Phi_2& =&sy_0\Bigg\{(T+\tau)(y+1)+\frac{(p_{\perp}-a)^2}{y_0^2}\left(1+\frac{\tau}{T}\right)
+\left(1+\frac{1}{y}\right)\frac{m^2}{y_0^2}\nonumber\\
&&+\frac{(\omega s)^2}{3y_0^2}\left[\frac{a^2}{Ty}+ap_{\perp}\left(1+\frac{\tau}{T}\right)\right]\Bigg\}\,.
\end{eqnarray}
Then, the integral over the variable $s$ can be taken via the saddle-point method. The saddle point is
\begin{eqnarray}\label{AB}
s_0&=&\frac{i}{\omega}\sqrt{\frac{A}{B}}\, ,\quad A=(T+\tau)(y+1)+\frac{(p_{\perp}-a)^2}{y_0^2}\left(1+\frac{\tau}{T}\right)
+\left(1+\frac{1}{y}\right)\frac{m^2}{y_0^2}\nonumber\\
&&B=\frac{1}{y_0^2}\left[\frac{a^2}{Ty}+ap_{\perp}\left(1+\frac{\tau}{T}\right)\right]\,.
\end{eqnarray}
In the vicinity of $s_0$ , the quantity $i\Phi_2$ reads
\begin{eqnarray}
i\Phi_2=-\left[\Phi_3+y_0\omega\sqrt{AB}(s-s_0)^2\right]\,,\quad \Phi_3=\frac{2y_0A}{3\omega}\sqrt{\frac{A}{B}}\, .
\end{eqnarray}
Now, it follows from Eq. (\ref{AB}) that the main contribution to the rate comes from the region of parameters
$$ \tau\ll 1\,,\quad p_{\perp}\approx a\,,\quad y\approx 1\,,\quad \frac{y_0}{m}\approx \frac{1}{\sqrt{2}} \,.$$
In this region, the phase $\Phi_3$ reads
\begin{eqnarray}
\Phi_3&=&\frac{1}{\chi}\Bigg[ 2\sqrt{3}(1+\tau)+\frac{\sqrt{3}(p_{\perp}-a)^2}{m^2}+\frac{8}{\sqrt{3}}\left(\frac{y_0}{m}-\frac{1}{\sqrt{2}}\right)^2\nonumber\\
&&+4\sqrt{\frac{2}{3}}\left(\frac{y_0}{m}-\frac{1}{\sqrt{2}}\right)(y-1)+\dfrac{7}{4\sqrt{3}}(y-1)^2\Bigg]\,,
\end{eqnarray}
and also $\omega|s_0|\approx \sqrt{3/2}m/a\ll 1$ and $|\rho|\approx 2\sqrt{3}a/\omega\gg 1$, as anticipated above. At this point we can evaluate the functions $X_1$ and $\bm Y_1$. For the evaluation of $X_1$ it is sufficient to set in the pre-exponent $s=s_0$, $y=1$, $\tau=0$, $p_{\perp}=a$, and $y_0/m=1/\sqrt{2}$ (except, of course, the overall factor $\tau$, see Eq. (\ref{eq:general6})). Instead, in order to evaluate the quantity $\bm Y_1$, it is also necessary to expand the pre-exponent in the vicinity of these points, taking into account strong compensation between different terms. After that we can take the integrals over $s$, $\tau$ and $y$. As a result, we finally obtain 
\begin{eqnarray}\label{eq:qq}
d\dot{W}&=&\frac{3^{1/4}}{48\sqrt{7}\pi^{5/2}}\frac{(Z\alpha)^2 }{ma}\chi^{3/2}\,dy_0\,d\bm p_{\perp}
\nonumber\\
&&\times\exp\left\{-\frac{\sqrt{3}}{\chi}\left[ 2+\frac{8}{7}\left(\frac{y_0}{m}-\frac{1}{\sqrt{2}}\right)^2+
\frac{(p_{\perp}-m\xi)^2}{m^2}\right]\right\}(1+\bm\zeta\bm\zeta_f) \,,\nonumber\\
&&\bm\zeta_f=-\mu\left[\frac{\chi}{\sqrt{2}}+\sqrt{6}\frac{(p_{\perp}-m\xi)^2}{m^2}\right]\frac{\bm p_{\perp}}{p_{\perp}} \, ,
\end{eqnarray}
where $\bm\zeta_f$ is written in the leading order with respect to $\chi\ll 1$. Note that we have also assumed that $\chi\gg 1/\xi$ and that $(p_{\perp}-m\xi)^2/m^2\lesssim \chi$. 

The term independent of the spin in Eq. (\ref{eq:qq}) is in agreement with the result in Eq. (4.39) in \cite{KuchievPRA}. Also, by integrating with respect to $y_0$ and $\bm{p}_{\perp}$ and by summing over the projections of the electron spin, one obtains the total rate in agreement with previous results, see e.g. Ref. \cite{Milstein}. The rate depends on the two invariant parameters $\xi$ and $\chi$. The parameter $\xi$ determines the typical transverse momentum of the created electron, i.e. $p_{\perp}\approx m\xi$ (see also \cite{Yakovlev,MVG,KuchievPRA}). Besides, Eq. (\ref{epsi}) together with the fact that $y_0\approx m/\sqrt{2}$ indicates that the typical energy of the electron is $\epsilon\approx m\xi^2/\sqrt{2}$. On the other hand, the parameter $\chi$ determines the width of the energy distribution and of the transverse momentum distribution around the mentioned positions where the respective maximum lies. In the region of parameters under consideration these distributions are rather narrow because $\chi\ll 1$. 

Concerning the polarization, our result indicates that the average electron (positron) polarization $\bm\zeta_f$ is collinear with $\bm p_{\perp}$. As we have mentioned, in the Born approximation with respect to the Coulomb field $\bm\zeta_f$ is independent of the nuclear charge number $Z$. Also, though the pair production rate is exponentially small at $\chi\ll 1$, the polarization $\bm\zeta_f$ is a linear function of $\chi$ in the regime where $1/\xi\ll \chi\ll 1$. Such a dependence has also been found for pair production processes in other fields configurations \cite{Novak}. In order to illustrate the polarization degree of the created electron let us use laser parameters which are not extreme nowadays (optical laser intensity of $10^{20}\;\text{W/cm$^2$}$) and the gamma factor of protons $\gamma=7500$ which will be soon available at the Large Hadron Collider (LHC). In this case we obtain that the polarization amounts to $|\bm\zeta_f|=\chi/\sqrt{2}\approx 0.2$ at $p_{\perp}=m\xi$.

\section{Conclusion}

In this paper we have investigated the process of $e^+e^-$ pair production in Coulomb and laser fields and the dependence of the production probability on the spin of one of the created particles. The pair production probability has been obtained by taking into account exactly the laser field and in the leading order the Coulomb field. In the case of a monochromatic circularly polarized undercritical laser field a compact, analytical expression of the pair production rate, differential in the degrees of freedom of one of the particles, has been obtained in the quasiclassical approximation. Our result shows that the average polarization vector of the electron (as well as of the positron) is directed along the momentum of the particle perpendicular to the laser propagation direction. Electron (positron) polarization degrees of the order of a few percent are predicted at laser intensities already available at table-top laser facilities combined with the LHC proton beam.

\section*{Acknowledgments}
A. I. M. gratefully acknowledges the hospitality and the financial support he has received during his visit at Max-Planck-Institute for Nuclear Physics. The work was supported in part by the RFBR Grant No. 09-02-00024 and Grant No. 14.740.11.0082 of federal program ``personnel of innovational Russia''.

\end{document}